\documentstyle[11pt,psfig,twoside]{article}

\begin{document}

\newcommand{\dd}{\,{\rm d}}
\newcommand{\ie}{{\it i.e.},\,}
\newcommand{\etal}{{\it et al.\ }}
\newcommand{\eg}{{\it e.g.},\,}
\newcommand{\cf}{{\it cf.\ }}
\newcommand{\vs}{{\it vs.\ }}
\newcommand{\zdot}{\makebox[0pt][l]{.}}
\newcommand{\up}[1]{\ifmmode^{\rm #1}\else$^{\rm #1}$\fi}
\newcommand{\dn}[1]{\ifmmode_{\rm #1}\else$_{\rm #1}$\fi}
\newcommand{\upd}{\up{d}}
\newcommand{\uph}{\up{h}}
\newcommand{\upm}{\up{m}}
\newcommand{\ups}{\up{s}}
\newcommand{\arcd}{\ifmmode^{\circ}\else$^{\circ}$\fi}
\newcommand{\arcm}{\ifmmode{'}\else$'$\fi}
\newcommand{\arcs}{\ifmmode{''}\else$''$\fi}
\newcommand{\MS}{{\rm M}\ifmmode_{\odot}\else$_{\odot}$\fi}
\newcommand{\RS}{{\rm R}\ifmmode_{\odot}\else$_{\odot}$\fi}
\newcommand{\LS}{{\rm L}\ifmmode_{\odot}\else$_{\odot}$\fi}

\newcommand{\Abstract}[2]{{\footnotesize\begin{center}ABSTRACT\end{center}
\vspace{1mm}\par#1\par
\noindent
{~}{\it #2}}}

\newcommand{\TabCap}[2]{\begin{center}\parbox[t]{#1}{\begin{center}
  \small {\spaceskip 2pt plus 1pt minus 1pt T a b l e}
  \refstepcounter{table}\thetable \\[2mm]
  \footnotesize #2 \end{center}}\end{center}}

\newcommand{\TableSep}[2]{\begin{table}[p]\vspace{#1}
\TabCap{#2}\end{table}}

\newcommand{\TableFont}{\footnotesize}
\newcommand{\TableFontIt}{\ttit}
\newcommand{\SetTableFont}[1]{\renewcommand{\TableFont}{#1}}

\newcommand{\MakeTable}[4]{\begin{table}[htb]\TabCap{#2}{#3}
  \begin{center} \TableFont \begin{tabular}{#1} #4 
  \end{tabular}\end{center}\end{table}}

\newcommand{\MakeTableSep}[4]{\begin{table}[p]\TabCap{#2}{#3}
  \begin{center} \TableFont \begin{tabular}{#1} #4 
  \end{tabular}\end{center}\end{table}}

\newenvironment{references}%
{
\footnotesize \frenchspacing
\renewcommand{\thesection}{}
\renewcommand{\in}{{\rm in }}
\renewcommand{\AA}{Astron.\ Astrophys.}
\newcommand{\AAS}{Astron.~Astrophys.~Suppl.~Ser.}
\newcommand{\ApJ}{Astrophys.\ J.}
\newcommand{\ApJS}{Astrophys.\ J.~Suppl.~Ser.}
\newcommand{\ApJL}{Astrophys.\ J.~Letters}
\newcommand{\AJ}{Astron.\ J.}
\newcommand{\IBVS}{IBVS}
\newcommand{\PASP}{P.A.S.P.}
\newcommand{\Acta}{Acta Astron.}
\newcommand{\MNRAS}{MNRAS}
\renewcommand{\and}{{\rm and }}
\section{{\rm REFERENCES}}
\sloppy \hyphenpenalty10000
\begin{list}{}{\leftmargin1cm\listparindent-1cm
\itemindent\listparindent\parsep0pt\itemsep0pt}}%
{\end{list}\vspace{2mm}}

\def\TYLDA{~}
\newlength{\DW}
\settowidth{\DW}{0}
\newcommand{\dw}{\hspace{\DW}}

\newcommand{\refitem}[5]{\item[]{#1} #2%
\def\REFARG{#3}\ifx\REFARG\TYLDA\else, {\it#3}\fi
\def\REFARG{#4}\ifx\REFARG\TYLDA\else, {\bf#4}\fi
\def\REFARG{#5}\ifx\REFARG\TYLDA\else, {#5}\fi.}

\newcommand{\Section}[1]{\section{#1}}
\newcommand{\Subsection}[1]{\subsection{#1}}
\newcommand{\Acknow}[1]{\par\vspace{5mm}{\bf Acknowledgements.} #1}
\pagestyle{myheadings}

\def\thefootnote{\fnsymbol{footnote}}
\begin{center}
{\Large\bf The Optical Gravitational Lensing Experiment.\\
\vskip3pt
Photometry of the MACHO-SMC-98-1 Binary Microlensing Event
\footnote{Based on  observations obtained with the 1.3~m Warsaw
telescope at the Las Campanas  Observatory of the Carnegie Institution
of Washington.}}
\vskip1cm
{\bf A.~~U~d~a~l~s~k~i$^1$,~~M.~~K~u~b~i~a~k$^1$,~~M.~~S~z~y~m~a~{\'n}~s~k~i$^1$,\\ 
G.~~P~i~e~t~r~z~y~\'n~s~k~i$^1$,~~ 
P.~~W~o~\'z~n~i~a~k$^2$,~~ and~~K.~~\.Z~e~b~r~u~\'n$^1$}
\vskip3mm
{$^1$Warsaw University Observatory, Al.~Ujazdowskie~4, 00-478~Warszawa, Poland\\
e-mail: (udalski,mk,msz,pietrzyn,zebrun)@sirius.astrouw.edu.pl\\
$^2$ Princeton University Observatory, Princeton, NJ 08544-1001, USA\\
e-mail: wozniak@astro.princeton.edu}
\vskip1cm
\end{center}

\Abstract{
We present photometry of the unique binary microlensing event
MACHO-SMC-98-1  collected by the OGLE group. Particularly interesting
observation was made close to the first caustic crossing which was
not covered by observations of other groups. It allows to test proposed
models of which Model~1 proposed by  PLANET group  seems to be in
the best agreement with the OGLE observations.}

\Section{Introduction}

The unique binary microlensing event toward the Small Magellanic Cloud,
MACHO-SMC-98-1, was discovered on May~25, 1998 with the MACHO team alert
system (Pratt \etal 1996). On June~8, 1998 the MACHO team issued a
second level alert based on significant deviations of the observed light
magnification from regular, point source microlensing light curve (Becker
\etal 1998). Observed light variations were interpreted as caused by
binary microlensing and the MACHO group predicted the second caustic
crossing around June~20, 1998. The object was then extensively observed
by microlensing follow-up projects: PLANET (Albrow \etal 1998) and
MACHO/GMAN (Alcock \etal 1998). Refined fits to the observed light curves
led to more precise prediction of the second caustic crossing which took
place on June 18.2~UT, 1998. The second caustic crossing was very well
sampled by the PLANET team. Also its falling part was well covered by
the EROS group (Afonso \etal 1998).

Non-standard microlensing events, in particular binary events, provide a
rare opportunity to, at least partially, remove degeneracy between
geometry of the event and transverse velocity of the lens. This
degeneracy causes usually non-uniqueness of the regular event solution.
In the case of the binary microlensing event it is possible to measure
the time of the caustic crossing. This gives a possibility to measure
the proper motion of the lens relative to the amplified source. Such a
measurement immediately provides a constraint on the linear velocity as
a function of distance. In the case of the MACHO-SMC-98-01 event this
possibility was especially interesting as it made it possible to
distinguish the likely location of the lens, either in the Galactic halo
or the SMC itself, by straight comparison of the lens proper motion with
kinematics of the regular SMC or halo objects.

Detailed analyses of the MACHO-SMC-98-01 event have been already
published by PLANET (Albrow \etal 1998) and MACHO groups (Alcock \etal
1998). The analysis by PLANET was based on  modeling of the light curve
of the event collected after the first caustic crossing while the one by
MACHO -- on the light curve of the entire event. The PLANET group found
two acceptable models of the event reproducing well observed brightness 
(\cf Fig.~2, Albrow \etal 1998), giving, however,  different proper
motion values: 1.26 km/s/kpc -- Model~1 and 2.00 km/s/kpc -- Model~2.
The MACHO group model which also included EROS group observations was
claimed to be incompatible with both PLANET models and predicted the
mean transverse velocity of the lens $V=84\pm9$ km/s at the distance of
60~kpc (corresponding to the proper motion equal to 1.4 km/s/kpc). In
spite of these ambiguities of the binary lens solution the general
conclusion of both analyses was similar: the lensing object is located
in the SMC rather than in the Galactic halo.

MACHO-SMC-98-1 is located in the field SMC$\_$SC4 which is monitored
regularly during  the Optical Gravitational Lensing Experiment (OGLE)
(Udalski \etal 1998). Unfortunately, the lensed star turned out to be a
bit below detection limit in the regular 125 sec exposure {\it I}-band
images collected during the survey. Nevertheless, it showed up on some
images taken during the event when the star was magnified. What is more
important, one of observations was collected on June 6, when the first
large brightening due to crossing of the first caustic occurred.
Although the parts of the light curve of MACHO SMC-98-1 between
caustics, second caustic crossing and part of the light curve after the
second caustic crossing were well observed by different groups, the
region of high amplification around the first caustic was missed.
Because the caustic crossing regions are the most important parts of the
light curve of the binary microlensing event for distinguishing between
models, the OGLE observation might provide an important constraint for
modeling the event. In this paper we present photometry of
MACHO-SMC-98-1 collected during the event.

\Section{Observations}

All observations presented in this paper were obtained with the 1.3-m
Warsaw telescope equipped with the "first generation" CCD camera
(Udalski, Kubiak and Szyma{\'n}ski 1997) at the Las Campanas
Observatory, Chile, which is operated by the Carnegie Institution of
Washington.

\MakeTable{ccccc}{12.5cm}{OGLE observations of MACHO-SMC-98-1}
{\hline
\noalign{\vskip2pt}
Date& JD hel. & {\it I}-band mag & error & seeing\\
UT &-2450000 & & & \\
\noalign{\vskip2pt}
\hline
\noalign{\vskip2pt}
June ~2.4242 & 966.92556  & 20.35 & 0.25 & 1\zdot\arcs0\\
June ~6.4005 & 970.90195  & 18.38 & 0.05 & 1\zdot\arcs4\\
June 12.3809 & 976.88259  & 18.99 & 0.10 & 1\zdot\arcs6\\
June 22.3350 & 986.83693  & 19.38 & 0.13 & 1\zdot\arcs7\\
June 23.3486 & 987.85057  & 18.96 & 0.09 & 1\zdot\arcs3\\
June 27.4165 & 991.91859  & 19.52 & 0.14 & 1\zdot\arcs4\\
July ~2.4203 & 996.92250  & 20.00 & 0.22 & 1\zdot\arcs1\\
\noalign{\vskip2pt}
\hline
}

Due to poor weather conditions between May 29 and July 4, 1998 OGLE
collected only 9 {\it I}-band images of the relevant SMC$\_$SC4 field. 
All observations were reduced with the standard OGLE data pipeline
(Udalski, Kubiak and Szyma{\'n}ski 1997).

As mentioned in the Introduction the MACHO-SMC-98-1 star was above
detection limit only on a few frames. Table~1 provides a list of frames
on which the star was measurable. Photometry of MACHO-SMC-98-1 was
derived with the {\sc DoPhot} photometry program running on  $200\times
400$ pixel subframes around MACHO-SMC-98-1.  First, differential
photometry with respect to  three nearby constant stars was calculated.
Then  the absolute photometry was determined  using photometry of
comparison stars from {\it BVI} maps of the SMC (Udalski \etal 1998).
Magnitudes of MACHO-SMC-98-1 with their errors are listed in Table~1.

Fig.~1 presents a region around MACHO-SMC-98-1 from the image taken on
June 6, 1998 when the magnification was high. North is up and East to
the left. A few non-variable stars suitable for comparison with
observations collected with other instruments are marked. Standard band
{\it BVI} photometry (Udalski \etal 1998) of these stars is listed in
Table~2. Stars A, C and D were used for determination of magnitudes of
MACHO-SMC-98-1 listed in Table~1.

\MakeTable{cccc}{12.5cm}{Comparison stars for MACHO-SMC-98-1}
{\hline
\noalign{\vskip2pt}
Star& $B$ & $V$  & $I$\\
\noalign{\vskip2pt}
\hline
\noalign{\vskip2pt}
A & $17.974\pm0.021$ & $16.853\pm0.014$& $15.691\pm0.015$\\
B & $18.156\pm0.029$ & $16.894\pm0.018$& $15.730\pm0.015$\\
C & $18.790\pm0.045$ & $17.520\pm0.021$& $16.199\pm0.018$\\
D & $19.420\pm0.046$ & $18.507\pm0.018$& $17.423\pm0.024$\\
E & $19.738\pm0.050$ & $18.990\pm0.034$& $18.124\pm0.038$\\
\noalign{\vskip2pt}
\hline
}

OGLE fields are also monitored from time to time in the {\it B} and {\it
V} bands. MACHO-SMC-98-1 was found to be a bit above the detection
threshold in the {\it V}-band. 24 {\it V}-band observations of
MACHO-SMC-98-1 were collected between July 16, 1997 and January~5, 1998.
The mean {\it V}-band magnitude of the star was found to be
$V=21.41\pm0.23$. Both -- PLANET and MACHO models suggest that the
lensed object is blended with an optical counterpart. Therefore the
above value represents the {\it V}-band magnitude of the entire blend
rather than the baseline magnitude of the lensed object. One should be
also  aware that due to long time scale of the event (Albrow \etal 1998,
Alcock \etal 1998) the lensed star could have been slightly magnified
during the OGLE {\it V}-band measurements.

\Section{Discussion}
 
Fig.~2 presents OGLE observations of the MACHO-SMC-98-1 event. Solid and
dotted lines correspond to the theoretical light curves of binary event
proposed by PLANET collaboration which were kindly provided by Drs P.\
Sackett and M.\ Dominik. As the PLANET group observations were made in
the {\it I}-band and were calibrated based on the OGLE maps of the SMC,
the comparison of their results with OGLE photometry is straightforward.

Fig.~3 presents enlargement of the first caustic crossing region. As can
be seen the PLANET Model~1 light curve (solid line) is in very good
agreement with the observed brightness of MACHO-SMC-98-1 at that time.
PLANET Model~2 predicts earlier caustic crossing and therefore seems to
be incompatible with observations. Also the remaining OGLE observations
are in good agreement with PLANET Model~1. The largest deviation is
equal to $1.5\sigma$, for the observation taken before the first caustic
crossing and it is possible that the PLANET Model~1 requires some fine
tuning. Also observation taken on June~22 deviates by about 0.25 mag
from the Model~1 light curve and PLANET observations taken earlier that
night. Although it  was obtained at worse seeing conditions (about 1.7
arcsec) we did not find any reason responsible for so large deviation.
Fortunately the star was observed almost at the same time by the
MACHO/GMAN  group and their {\it R}-band data from CTIO are available
from their WEB site {\it http://darkstar.astro.washington.edu}. Two
observations were collected by the MACHO/GMAN  group on June 22. The
first one was made about 0.0308~day after the OGLE observation and
is about 0.25~mag fainter than the next one taken another 0.0195~day
later. Therefore we are convinced that the observed drop of the
MACHO-SMC-98-1 brightness by about 0.25~mag on June~22 is real.

Unfortunately it is not simple to compare our data with MACHO/GMAN
model. The data they analyzed were taken in different, usually
non-standard bands making direct comparison difficult without good
coverage of the entire light curve. We approached this problem in the
following way. First, we realized that the observation taken on June~12,
1998 during the flat plateau between caustic crossings was collected at
the same time as the standard {\it R}-band observation obtained by
MACHO/GMAN at CTIO. The amplification according to MACHO/GMAN was at
that moment equal to $A=11.25\pm0.1$ (see Fig.~2, Alcock \etal 1998). If
we assume that the event was monochromatic, and the contribution of the
lensed star to the total brightness of the object equal to 0.6 in the
{\it I}-band (conservative assumption based on MACHO/GMAN data for CTIO
{\it B} and {\it R}-band observations -- Table~2, Alcock \etal 1998),
then we may calculate amplification on June 6, 1998 \ie close to the
first caustic crossing, based on our observed magnitude differences
between June 6 and June 12 measurements. The resulting amplification on
June 6.4005 UT (JD hel.=2450970.90192) was $A=20.2\pm1.1$ where the
error includes uncertainties in magnitude differences and uncertainty of
June~12 amplification. It should be noted that the OGLE observation from
June~12, serving as the base of amplification calculations, is in
excellent agreement with measurements obtained by PLANET group during
that night.

Fig.~4 presents the MACHO model light curve close to the first caustic
crossing kindly provided by Dr.\ D.\ Bennett together with the OGLE
observation of June~6.  It is clear that the MACHO/GMAN model prediction
of the first caustic crossing is delayed by about 0.14~day and predicted
amplification for the epoch of the OGLE  observation is more than
$3\sigma$ larger than that resulting from  observation.

Concluding, it seems that the PLANET group Model~1 provides the best
agreement with the OGLE observation collected close to the first caustic
crossing and the remaining observations.  It may, however,  require some
fine tuning as the observation collected before the first caustic
crossing deviates by about $1.5\sigma$ from predicted light curve.
PLANET Model~2 and MACHO/GMAN model predict the first caustic crossing
too early and too late, respectively. As the Model~1 of the PLANET group
predicts the slowest proper motion of the lensing object, the general
conclusion reached by all groups that the MACHO-SMC-98-1 lens was
located in the SMC is strongly supported.

{\bf Acknowledgements.} We would like to thank Drs.\ P.\ Sackett and M.\
Dominik from the PLANET group and Dr.\ D.\ Bennett from MACHO group for
providing us with their model light curves of the MACHO-SMC-98-1 event.
We thank Prof.\ B.\ Paczy{\'n}ski for discussions and comments. The 
paper was partly supported by the Polish KBN grant 2P03D00814 to A.\
Udalski.  Partial support for the OGLE project was provided with the NSF
grant AST-9530478 to B.~Paczy\'nski.

\newpage

\centerline{\large\bf Figure Captions}
\vskip1cm 
\noindent 
Fig.~1. $83\arcs\times 166\arcs$ {\it I}-band subframe around
MACHO-SMC-98-1 taken on June 6.4005 UT, 1998 close to the first caustic
crossing. North is up and East to the left. Positions of comparison
stars are marked by letters 'A'--'E'. 'M' denotes position of
MACHO-SMC-98-1.

\noindent
Fig.~2. OGLE observations of MACHO-SMC-98-1. Solid and dotted lines
represent the PLANET group Model~1 and Model~2 light curves, respectively. 

\noindent
Fig.~3. Enlargement of Fig.~2 around the first caustic crossing.

\noindent
Fig.~4. The MACHO group model light curve of MACHO-1998-SMC-1 with OGLE 
observation.

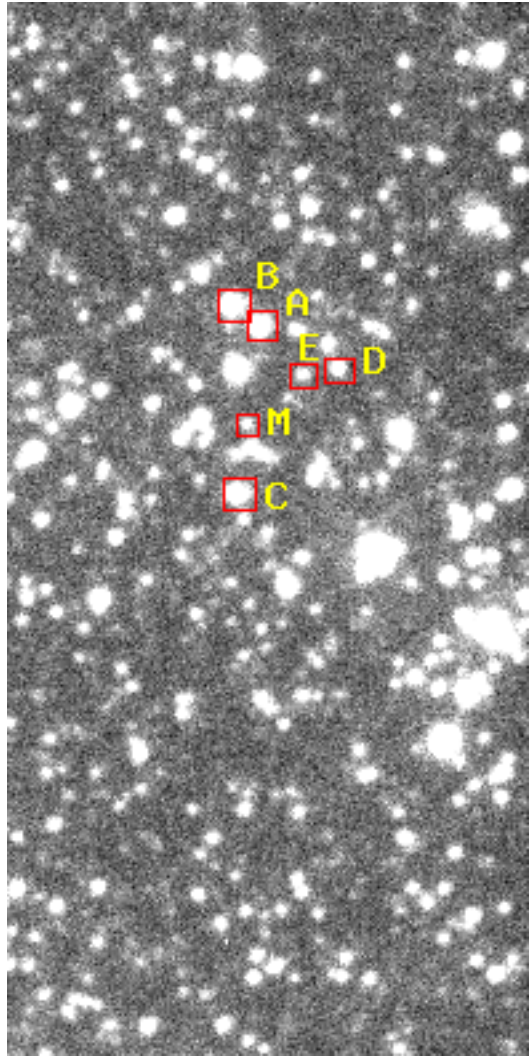
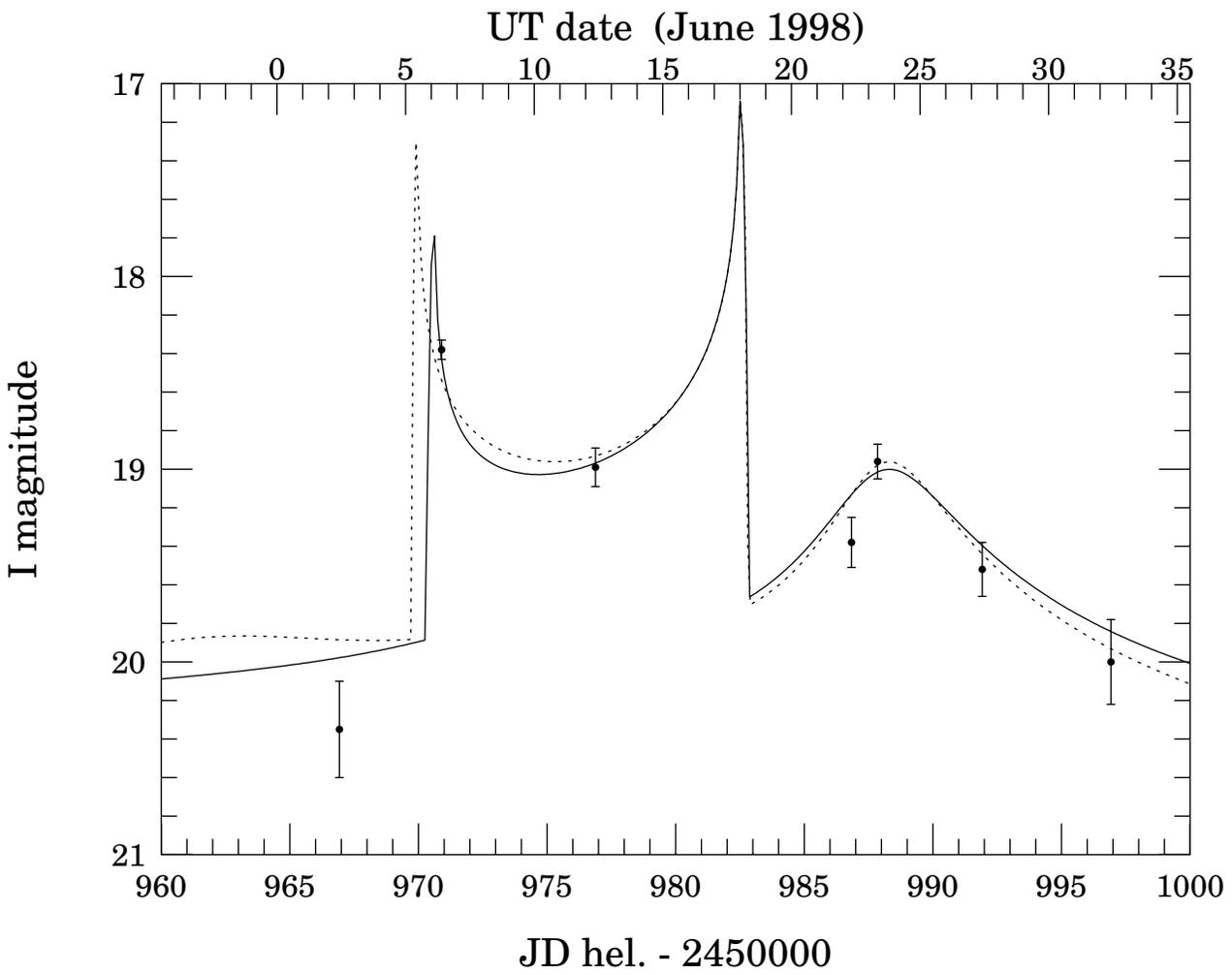
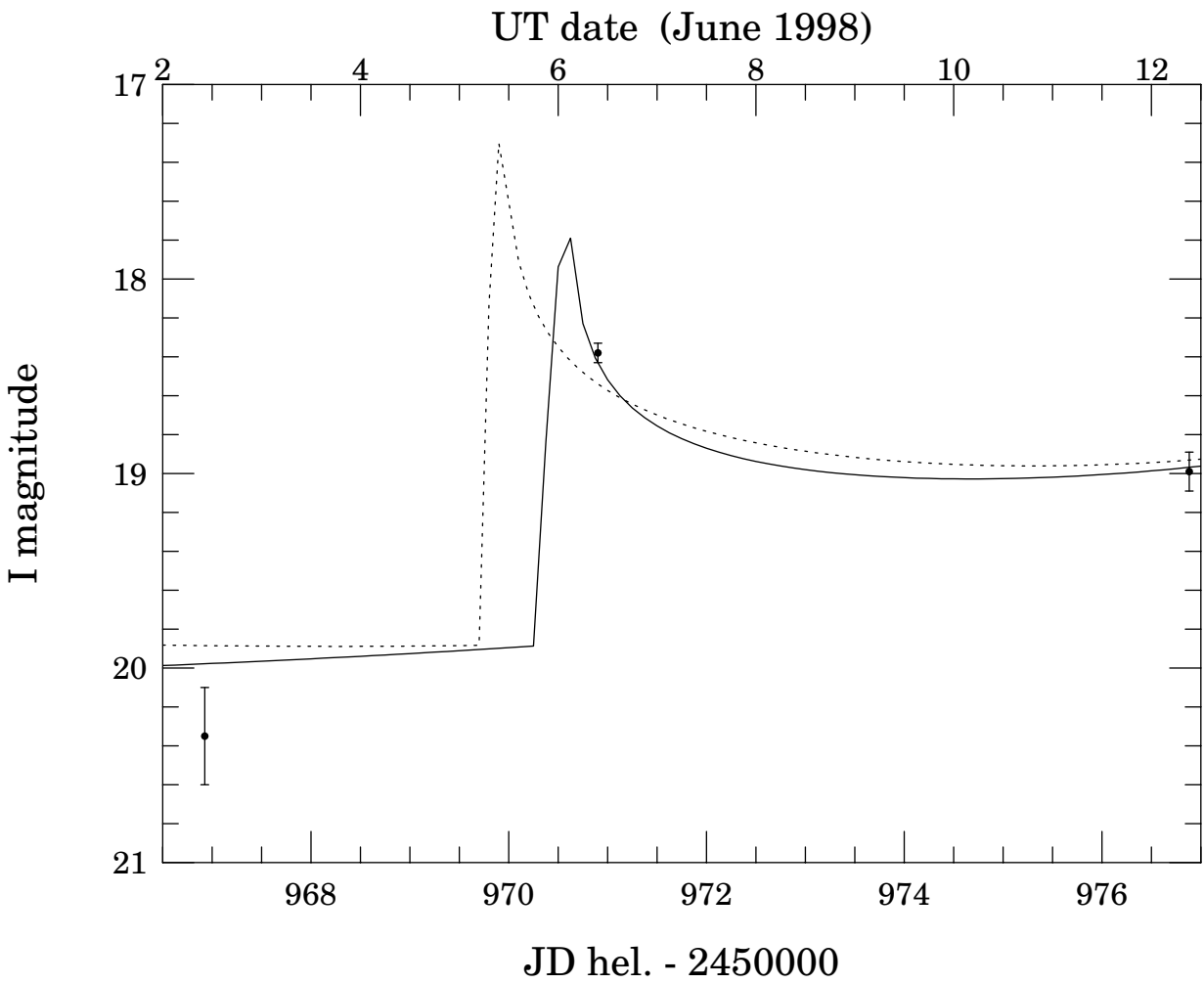
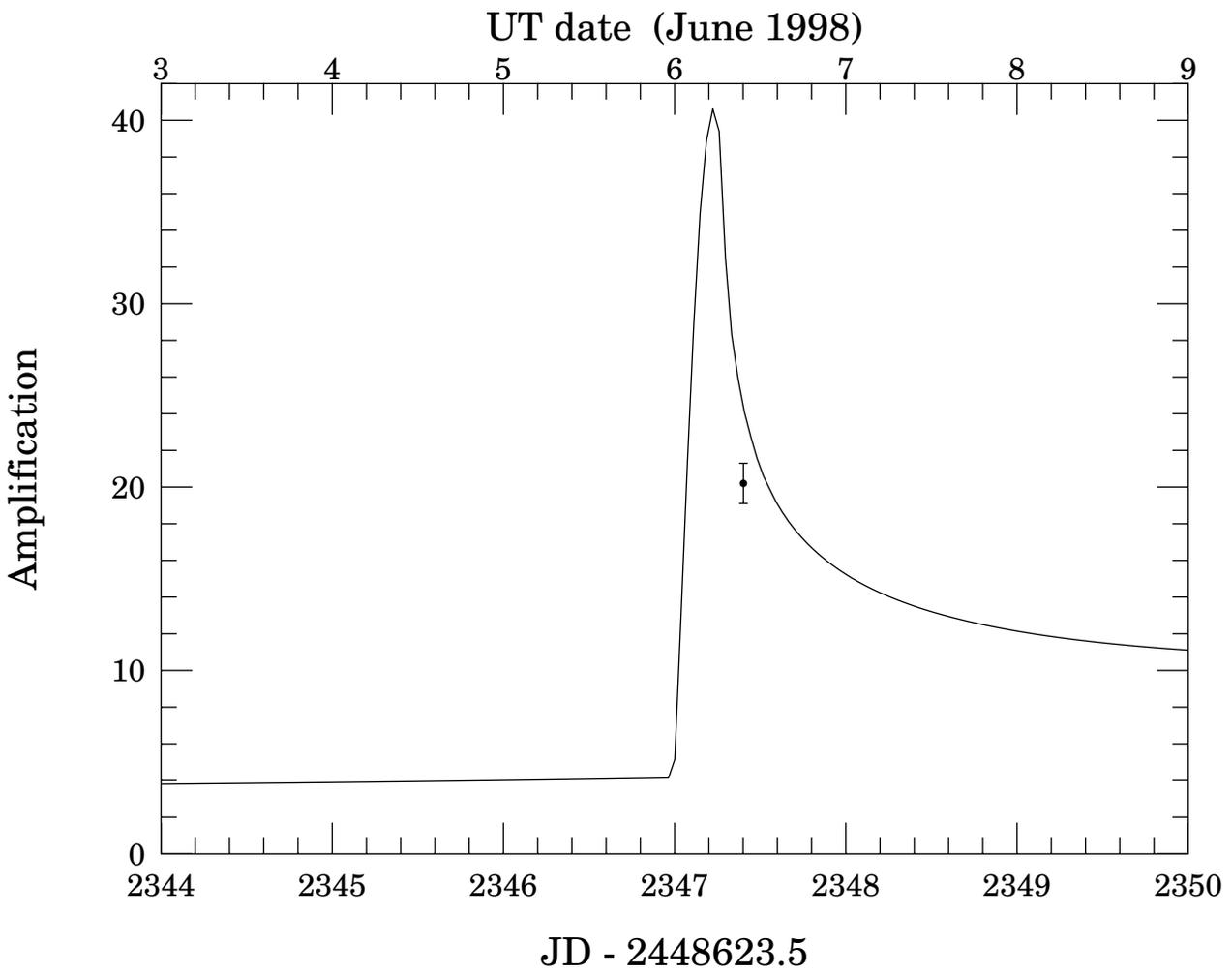
\end{document}